\documentclass[prc,preprint]{revtex4} 
\usepackage{graphicx}

\newcommand{\be}{\begin{eqnarray}}
\newcommand{\ee}{\end{eqnarray}}

\begin{document}
%\twocolumn[\hsize\textwidth\columnwidth\hsize\csname @twocolumnfalse\endcsname

\title {
Chemical Freezeout in Heavy Ion Collisions
}
\author{Derek Teaney}
\affiliation {
 Department of Physics, Brookhaven National Laboratory,
     Upton, NY 11973-5000
}
\date{\today}
\begin{abstract}
We construct a hadronic equation of state consistent with chemical
freezeout and discuss how such an equation of state modifies the
radial and elliptic flow in a hydrodynamic + hadronic cascade model 
of relativistic heavy ion collisions at the SPS.
Incorporating chemical freezeout does not change the 
relation between pressure and energy density.
However, it does change the relation between temperature and
energy density. 
Consequently, when
the hydrodynamic solution and freezeout are expressed
in terms of energy density,  chemical freezeout does not
modify the hydrodynamic radial and elliptic flow velocities studied previously.
Finally, we examine 
chemical freezeout within the hadronic cascade (RQMD). 
Once chemical freezeout is incorporated into the hydrodynamics, the
final spectra and fireball lifetimes are insensitive to 
the temperature at which the switch from hydrodynamics 
to cascade is made. 
Closer inspection indicates that the pion spectrum 
in chemically frozen hydrodynamics is significantly cooler than in
the hydro+cascade model. This difference is reflected in $v_{2}(p_{T})$.
We extract the freezeout hadron density in
RQMD and interpret it in thermal terms; the freezeout hadron 
density corresponds to a freezeout temperature of 
$T_{f}\approx100\,$ MeV 
and $\mu_{\pi} \approx 80\,$ MeV.

\end{abstract}
\maketitle
\vspace{0.1in}
\newpage

\section{Introduction}
Currently, by colliding heavy ions at the
Super Proton Synchrotron (SPS) and the Relativistic Heavy Ion Collider (RHIC), 
experimentalists \cite{QuarkMatter99,QM2001} have endeavored to  create
a deconfined state of quarks and gluons -- the Quark Gluon Plasma 
(QGP).  
Some degree of thermalization in the heavy ion reaction is a prerequisite
for QGP formation.
It is an experimental fact that
the ratios of the twenty or so hadron species produced in the heavy ion
collision are close to the thermal ratios expected
of an ideal gas of hadrons at a temperature of $T\approx160-180\,\mbox{MeV}$ 
\cite{Thermal}.
It is exciting that this temperature is close to the transition
temperature to the QGP, $T_{c}\approx160\,\mbox{MeV}$.
However, after accounting for finite size
corrections, the hadron ratios in $p\bar{p}$ and $e^{+}e^{-}$
reactions are
reproduced by the same thermal models used to
describe heavy ion data. It seems
that statistical mechanics provides a universal description of
hadronization. Unlike $p\bar{p}$ and $e^{+}e^{-}$ reactions,
the $\sim5000$
hadrons produced in the heavy ion collision
rescatter after hadronization. Ideally, this ensemble of hadrons
may be considered a hot hadronic gas; hydrodynamics then
describes the subsequent evolution. With an understanding of
the final hadronic expansion, the global properties of the collision can
be quantified.

We first review the notion of chemical freezeout in the 
hadron gas  \cite{Bebie-Chemical1,Ed-Chemical,Leutwyler-Chemical1,Gavin-Chemical}.
In the hadron gas,
the principal hadronic reactions, e.g.
$\pi\pi\rightarrow\rho\rightarrow\pi\pi$, $\pi K\rightarrow
K^{*}\rightarrow\pi K$, $\pi N\rightarrow\Delta\rightarrow\pi N$,
do not change the yields of pions, kaons, and nucleons. 
Nevertheless, these reactions provide a mechanism
for thermalization during the hadronic evolution. Thermal
equilibration times at a temperature of $T\approx160\,\mbox{MeV}$
are typically $\sim 2\,\mbox{fm/c}$ \cite{Raju-Nonequilibrium}.
In the strong interactions of light hadrons there are
only three conserved currents:
baryon number ($J_{B}^{\mu}$), strangeness
($J_{S}^{\mu}$), and isospin ($J_{I}^{\mu}$).
Other hadronic reactions (for instance
$\pi N\rightarrow N^{*}(1530)\rightarrow\Delta\pi\rightarrow N\pi\pi$)
do change the total yield of pions, kaons and nucleons.
Because of these reactions the system approaches
chemical equilibrium. The yield of pions, kaons, and nucleons changes via
such reactions until the
Gibbs free energy reaches a minimum. However, the time scale
of chemical equilibration is much longer than that of
thermal equilibration;
chemical equilibration times are
typically $\tau_{ch}\sim200\,\mbox{fm/c}$  
\cite{Song-Chemical,Pratt-Chemical}.
Therefore in the hadron gas, there are two disparate time scales,
$\tau_{th}$ and $\tau_{ch}$.

In heavy ion collisions, the lifetime of the hadronic
stage is approximately $\sim 10\,\mbox{fm/c}$, which is
very short compared to $\tau_{ch}$ , but longer
than $\tau_{th}$.
Therefore, on the time scale
of the collision, although the chemical composition is
fixed at the time of hadronization, the system continues
to evolve kinetically for some time until the particles
breakup. The stages of the collision have been
described with the following picture. First,
there is chemical freezeout (hadronization). Then the
hadrons evolve as a hadronic fluid until thermal freezeout (breakup).
In this picture,
the particles develop chemical potentials during the hadronic
evolution since the total number of particles is
fixed and the temperature decreases.

There is some evidence for this chemical/thermal freezeout
picture. First,
hadronic cascades indicate that following hadronization
pions and nucleons rescatter for $\sim10$ fm/c. Consequently, the
pion $\langle p_T \rangle$ decreases while the nucleon
$\langle p_T \rangle$ increases.
Thus, microscopic calculations indicate that pions
cool and increase the transverse flow of the nucleons  \cite{HydroUrqmd}.
Second,
Rapp and Shuryak have argued that a pion chemical potential can
explain the anti-baryon yield at the SPS, since
$p$ and $\bar{p}$ can be produced via the forward and
backward reactions of $\bar{p}p\leftrightarrow 5\pi$  \cite{Rapp-Chemical}.
Finally, the most compelling experimental evidence for
the pion chemical potential comes from a combined analysis of
spectra and HBT correlations  \cite{Ferenc-UniversalF}. This study indicates that the hadron density
at freezeout is substantially lower than the density
of a hadron gas at $T\approx160\,\mbox{MeV}$. Furthermore, the
pion phase space density, which can be extracted from
from two pion correlations, is overpopulated.
An overpopulated phase space is expected from a Bose
gas with a chemical potential.

Hydrodynamics has been used extensively to model heavy ion
reactions.
However, these hydrodynamic calculations assume thermal
$and$ chemical equilibrium. While thermal equilibration is
at least plausible, chemical equilibration is certainly impossible.
Typically a hydrodynamic simulation is run until a universal
freezeout temperature $T_{f}\approx120-140\,\mbox{MeV}$, which
is adjusted to match the pion and nucleon spectra. Subsequently,
even if the spectral shape is correct, the
yields of $\bar{p}, \Lambda, K$ for example, are typically wrong
by factors of two. To account for this discrepancy,
a comparison to the data is then made
by rescaling the particle yields (by hand) to
their value at $T\approx160\,\mbox{MeV}$. Energy-momentum and
number conservation are violated in this
inconsistent procedure.

One approach to the problem of chemical freezeout is to
stop the hydrodynamic evolution at $T_{c}$ and then continue
the evolution with a hadronic cascade \cite{HydroUrqmd,Htoh}.
This approach incorporates chemical freezeout as chemical
equilibration times are encoded into the hadronic cross sections
in the cascade. Although hydro+cascade provides a comprehensive
description of the heavy ion data and has been
used extensively, the final spectra are
sensitive to exactly where the switch to the cascade is made 
\cite{HydroUrqmd}.
Ideally, the switching temperature, $T_{switch}$,
should be varied, with final results
independent of this artificial parameter.
However, the cascade conserves $\pi, K, \Lambda\dots$ number
in the dominant reactions; therefore to achieve
a kind of dual description chemical freezeout must
be incorporated into the hydrodynamic evolution.

This incorporation may be achieved by including additional
conservation laws for
$T<T_{c}$. For each conserved species $\pi, K,\Lambda\cdots$,
we have
\begin{eqnarray}
\label{EoM}
\partial_{\mu} J_{\pi}^{\mu} = \partial_{\mu} J_{K}^{\mu} =
\partial_{\mu} J_{\Lambda}^{\mu} = \cdots = 0.
\end{eqnarray}
The Equation of State(EOS) is modified and now depends on
$n_{\pi}, n_{K}, n_{\Lambda}\cdots $, in addition
to $\epsilon$ and $n_{B}$. Furthermore, the relationship
between energy density and temperature is dramatically different.
In this work, we consistently incorporate chemical freezeout
into the hydrodynamic evolution of the hydro+cascade model, 
H2H \cite{Htoh}. Similar recent efforts have been presented 
\cite{Grassi-Mu,Hirano-Mu}. 
H2H \cite{Htoh} and other hydrodynamic models 
\cite{Huovinen,Hirano-Flow}  
have been compared extensively to available data both at the
SPS and RHIC. The purpose of this work is to illustrate
how chemical freezeout changes the results  of these works. 

Briefly, in the H2H  model the initial stages of the 
collision are modeled with (2+1) ideal fluid dynamics, assuming
Bjorken scaling in the longitudinal direction 
\cite{Bjorken-83,Ollitrault-Elliptic}. 
The initial entropy is distributed in the 
transverse plane according to a Glauber model at a time,
 $\tau_{o} =1$ fm/c.
At particular temperature, $T_{switch}$ (which is 
less than the critical temperature $T_{c}=165\,\mbox{MeV}$), 
the fluid
is converted into hadrons via the Cooper-Frye formula \cite{CooperFrye}.
The Cooper-Frye formula is 
appended with a theta function rejecting backward moving hadrons.
Initial conditions for the hydrodynamic evolution have been 
chosen to match the model charged particle  multiplicity and  net baryon
number to the experimental values at the SPS (PbPb with 
$\sqrt{s}=17$ GeV A) and RHIC (AuAu with $\sqrt{s}=130$ GeV A) \cite{Htoh}.

The calculation presented here are for the SPS (PbPb with
$\sqrt{s}=17$ GeV A)  at an impact parameter of b=6 fm. (A non-central
impact parameter is taken in order to
study elliptic flow. Final flow velocities and freezeout conditions do 
not change rapidly with impact parameter \cite{Htoh,Kolb-Centrality}.)
The initial conditions (see Ref. \cite{Htoh} for more
detail) are  characterized by an
entropy per baryon ratio, $s/n_{B} = 42$, and  an average 
initial energy density, $\left.<\epsilon>\right|_{\tau_{o}} 
= 5.4$ $\mbox{GeV/fm}^3$. 
The EOS has an $0.8$ $\mbox{GeV/fm}^{3}$ latent heat and  
is identical (above $T_{c}$) to the equation of state LH8 which was 
used previously \cite{Htoh}. 
As this work concentrates on the freezeout stages of
the collision, the differences between the presented results and
the analogous calculations at RHIC are small \cite{Teaney-Thesis}. 
The slightly larger baryon density at the SPS does not significantly
alter the freezeout dynamics which is already meson dominated at 
the SPS \cite{Htoh,HydroUrqmd}

Sect.~\ref{MultiHydro} discusses
the EOS in which all the hadronic
species are conserved. The hydrodynamic
equations are then solved with and without the additional conservation
laws and the solutions contrasted. The principal result is
that the particle yields of a hydrodynamic calculation can
be consistently modified (increased), provided the freezeout temperature is
also modified (decreased), to keep the energy density the same.
Returning to the hydro+cascade approach, in Sect.~\ref{RQMD}, we
vary the switching temperature to the cascade, $T_{switch}$.
For $115\,\mbox{MeV} < T_{switch}< 160\,\mbox{MeV}$,
the model spectra are insensitive to variations.
Elliptic flow remains somewhat sensitive to the switching
temperature, although the sensitivity is reduced when
chemical freezeout is incorporated.

With these basic results, the freezeout conditions of
the cascade RQMD are analyzed in Sect.~\ref{RQMD}. At freezeout,
the density of pions in the cascade is equal
to the density of an ideal hadron gas with $T_{f}\approx100\,\mbox{MeV}$
and $\mu_{\pi}\approx80\,\mbox{MeV}$. This freezeout density
is universal and is
independent of impact parameter, collision energy
and the switching temperature, $T_{switch}$ \cite{Htoh,HydroUrqmd}. 
In evaluating the extent to
which the cascade reproduces the hydrodynamics in this low
temperature region, we find that
the hydrodynamics cools much more quickly than the cascade.
This rapid
cooling impacts both spectra and $v_{2}(p_{T})$.

\section{Implementing Chemical Freezeout in the EOS}
\label{MultiHydro}

In this section, we construct an EOS which incorporates
chemical freezeout. (For similar constructions see 
\cite{Bebie-Chemical1,Pratt-Chemical,Grassi-Mu,RappWambach}.) Above 
$T_c$, the EOS is identical to
LH8 which was described in \cite{Htoh}. Only
the hadronic portion of the EOS is modified.

For the purposes of this work, we consider expanding
hadronic gas with only the lowest hadron multiplets. Specifically,
we consider only the $0^{-}$ and $1^{-}$ meson octets, $\eta'$ and $\phi$,
the $\left. \frac{1}{2}\right.^{+}$ baryon and anti-baryon
octets and the $\left. \frac{3}{2}\right.^{+}$ baryon and
anti-baryon decuplets.
The strong resonant reactions in this ensemble excite
mesons and baryons from the lower hadronic multiplet to
the higher multiplet and are given (up to isospin and baryon/anti-baryon
symmetry) by
\begin{eqnarray}
\pi \pi \rightarrow \rho \rightarrow \pi\pi \\ \nonumber
\pi K^{+} \rightarrow K^{+*} \rightarrow \pi K^{+} \\ \nonumber
\pi K^{-} \rightarrow K^{-*} \rightarrow \pi K^{-} \\ \nonumber
\pi N \rightarrow \Delta \rightarrow \pi N \\ \nonumber
\pi Y \rightarrow \Sigma^{*} \rightarrow \pi Y \\ \nonumber
\pi \Xi \rightarrow \Xi^{*} \rightarrow \pi \Xi,
\end{eqnarray}
where Y denotes the hyperons $\Lambda,\Sigma$.
These are certainly the most important reactions for the
late stage of the heavy ion collision. If only these
reactions and elastic collisions are included, then
the system has 16 conserved currents: baryon number($J_{B}^{\mu}$),
strangeness($J_{S}^{\mu}$), isospin ($J_{I}^{\mu}$) and
13 other conserved numbers, $J_{H_i}^{\mu}$, where
$H_i$ runs over the hadron species,
\begin{equation}
\label{Hlist}
\pi,\bar{N},Y,\bar{Y},\Xi,\bar{\Xi},K^{-},\eta,\eta',
\omega,\phi,\Omega,\bar{\Omega}.
\end{equation}
Nucleon number and
$K^{+}$ number are not included in the list of hadrons, but their
conservation
follows from the conserved hadron currents already specified and from
baryon and strangeness conservation.
Perhaps some of the stranger species ($\Omega,\phi$)
should not be included in the list of hadrons
since they decouple early in the hadronic evolution.
In addition, $\omega$ does not neatly
fit into the thermal/chemical freezeout picture, as
its lifetime is comparable to the collision lifetime.
Nevertheless,
these particles are included for theoretical consistency and
we have found that they do not affect the bulk properties.

With the assumption of equilibrium the hydrodynamic
equations become
\begin{eqnarray}
\label{niequ}
\partial_{\mu} T^{\mu\nu} =0 \\
\partial_{\mu} ( n_{i} u^{\mu} ) = 0,
\end{eqnarray}
where $T^{\mu\nu} = (e + p) u^{\mu}u^{\nu} -p g^{\mu\nu}$.
This set of equations, together with the assumption of
smooth flow, imply entropy conservation,
\begin{eqnarray}
\label{sequ}
\partial_{\mu} (s u^{\mu}) = 0 .
\end{eqnarray}
An EOS -- a relation between the pressure and the
energy density and number densities associated
with the 16 conserved currents -- is required
to complete the system.

The hadronic EOS is taken as a sum of independent ideal
gases over the hadrons considered.
This approximation is based upon
the fact that the thermodynamics
of pions interacting via
$\pi\pi\rightarrow\rho\rightarrow\pi\pi$ is
nearly equivalent to the thermodynamics of an
ideal gas of $\pi$-s and $\rho$-s  \cite{Raju-ResonanceGas}.
In all the ideal gas formulas,
the 16
chemical potentials all appear in the
combination,
\begin{eqnarray}
\mu_{B} B + \mu_{S} S + \mu_{I} I + \sum_{i} \mu_{H_i} H_i ,
\end{eqnarray}
where $H_i$ runs over the hadron list given in Eq.~\ref{Hlist}.
For instance, $\pi^{-}$ has $I=-1/2 , H_{\pi}=1$, and
all other quantum numbers zero. On the other hand, $K^{-*}$
has $B=0$, $S=1$, $I=-1/2$, $H_{\pi}=1$, $H_{K^{-}}=1$ and
all other quantum numbers zero. The EOS is a relationship
between the pressure $p$, and the energy density $e$, and the
number densities $\{n_{i}\}$, and is therefore a
function of 1+16=17 variables.

In the heavy ion collision, only a small fraction of
the total 17 variable phase space is sampled.
Indeed, if at some time
$n_{i}/s$ is constant everywhere in space, then  Eq.~\ref{sequ}
and Eq.~\ref{niequ} imply that $n_{i}/s$ is a constant
for all later times \cite{Greiner,LL-HydroPg2}. (This follows directly since
$\partial_{\mu} \log(s) = \partial_{\mu} \log (n_{i}) =
-\partial_{\mu}u^{\mu}$ .)
Thus with the assumption of entropy conservation, it is
sufficient to know the EOS along
the trajectory where $n_{i}/s$ is constant. We now
specify $n_{i}/s$ and construct the hadronic EOS along
this trajectory. This unique trajectory in the
17 variable phase space is called the adiabatic path below.

The ratio $s/n_{B}=42$ is fixed by the
initial conditions and
is chosen to reproduce the experimental
proton to pion ratio \cite{Htoh}. The
net strangeness is zero, $n_{s}=0$. For simplicity, the net isospin is
taken to be
zero, $n_{I}=0$.
As discussed in the
introduction, at hadronization the system is
born into chemical equilibrium. Therefore at $T_{c}$,
$n_{H}/s$ is taken to be its chemically equilibrated value. Now,
with $n_{i}/s$ a specified constant on the $T=T_{c}$ hyper-surface,
as the system expands and cools, $n_{i}/s$ remains constant. The particles
develop chemical potentials to ensure this constancy. At a computational
level,
the procedure for constructing the EOS along the adiabatic trajectory is
the following. First, at $T=T_{c}$ adjust $\mu_{B}$,$\mu_{S}$, and
$\mu_{I}$, until $n_{S}=n_{I}=0$ and $s/n_{B}=42$.
Then, calculate $n_{H}/s$ at $T_{c}$. Then in small increments lower
the temperature, adjust all the chemical potentials to
leave $n_{i}/s$ constant, and tabulate all the thermodynamic quantities
(i.e, pressure, entropy, energy density) along the way.

\begin{figure}[tb!p]
\begin{center}
\includegraphics[height=3.0in,width=3.0in]{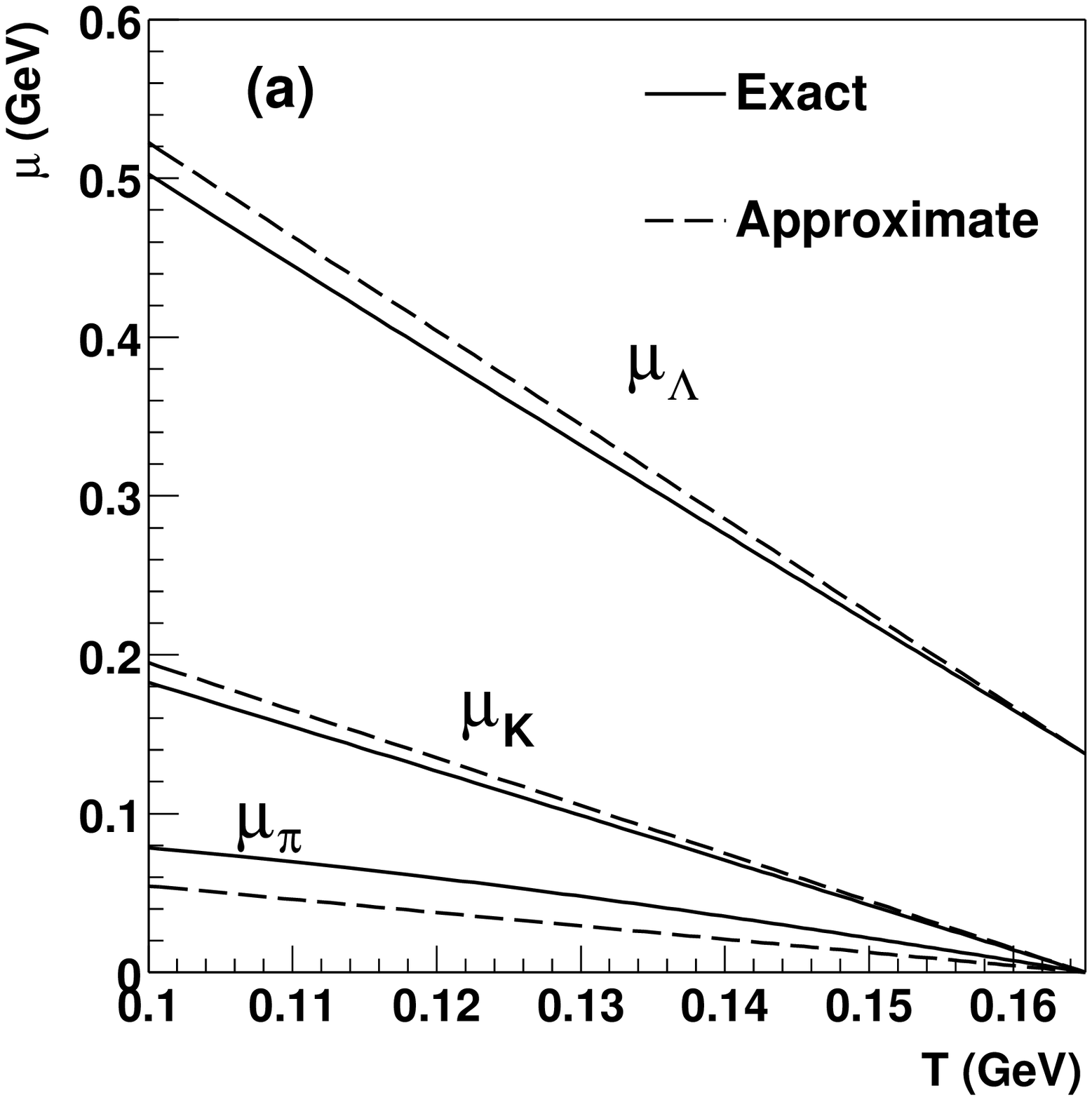}
\includegraphics[height=3.0in,width=3.0in]{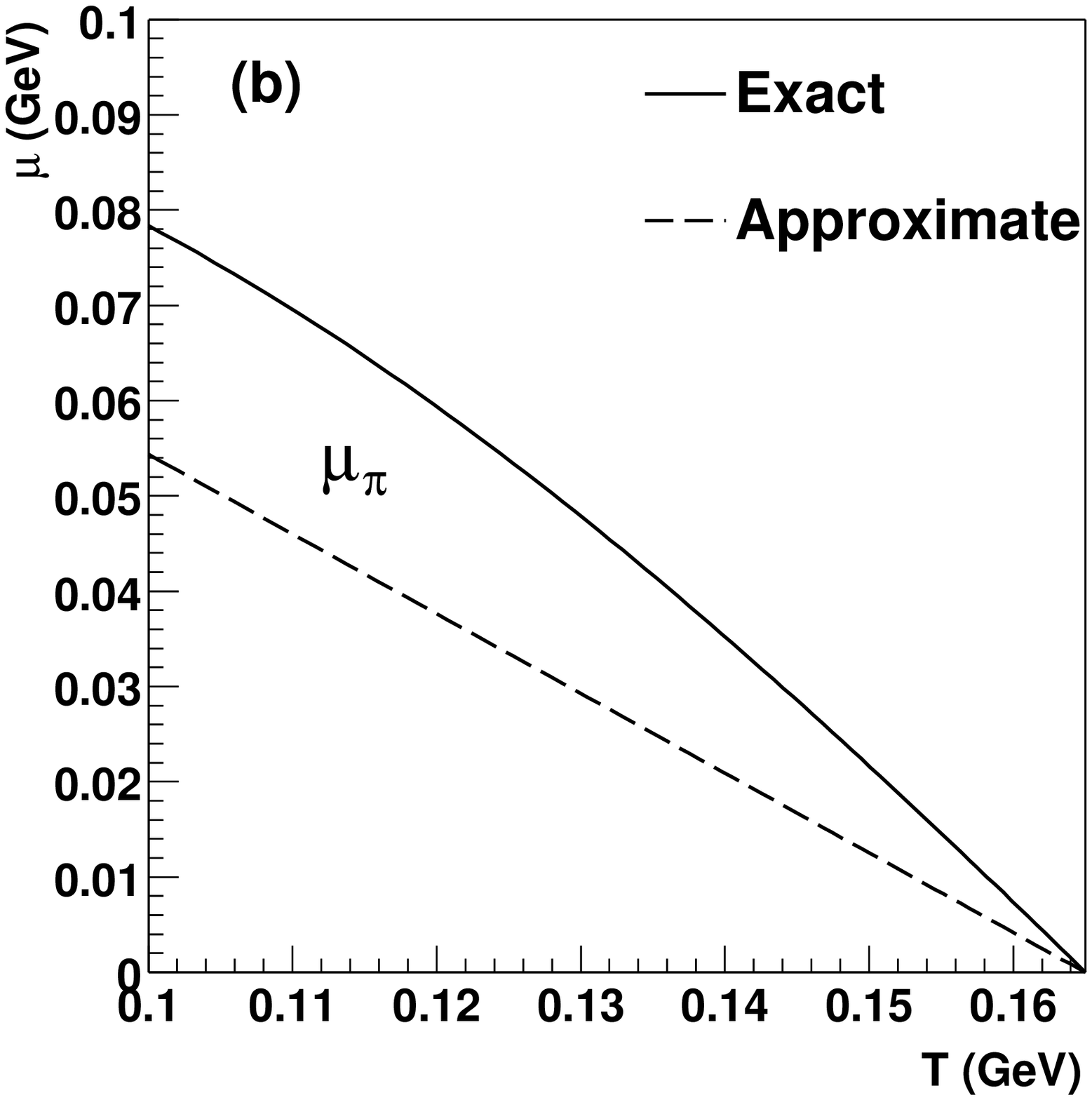}
\end{center}
\caption[Chemical potentials as a function of
temperature] {
\label{Tswitch}
Chemical potentials as a function
of temperature at the SPS ($s/n_{B}$=42) for (a) $\pi$, $K$ and $\Lambda$ and (b) $\pi$ only.
Eq.~\ref{MuFormula} gives an approximate formula
for these chemical potentials.
}
\end{figure}

Consider the results of this procedure. First,
a few of the chemical potentials
are shown as a function of the switching temperature
by the thick lines in Fig.~\ref{Tswitch}(a).
%%%%%%%%%%%%%%%%%%%%%%%%%%%%%%%%%%%%%%%%%%%%%%%%%%%%%%%%%%%%%%%%%%%%%%%%
%%%%%%%%%%%%%%%%%%%%%%%%%%%%%%%%%%%%%%%%%%%%%%%%%%%%%%%%%%%%%%%%%%%%%%%%
We can derive an approximate formula for the chemical
potentials as a function of $T$, for a collection of
non-relativistic ideal gases. For
non-degenerate and non-relativistic ideal gases the
partial pressure $p_{i}$, and partial energy density $e_i$
of the i-th species are given by,
\begin{eqnarray}
p_{i} &=& n_{i} T \\
e_{i} &=& n_{i} m_{i} + (\mbox{Const}) n_{i} T .
\end{eqnarray}
Since
$T\,s = \sum_{i} \epsilon_{i} + p_{i} - \mu_{i} n_{i} $ ,
\begin{eqnarray}
1 &=& \sum_{i} \frac{n_{i}}{s}
\left(\frac{m_{i} + \mu_{i}}{T} + \mbox{Const} \right).
\end{eqnarray}
$n_i/s$ is constant below for $T<T_{c}$. To keep all
the terms in parenthesis constant below $T_{c}$, we require
\begin{eqnarray}
\label{MuFormula}
\mu_{i} = m_{i} \left( 1- \frac{T}{T_c} \right) + \frac{T\mu_{c}}{T_c},
\end{eqnarray}
where $\mu_{c}$ is the chemical potential at the critical
temperature. The approximation is shown by the dashed lines and
works well for all particles except pions. Being the
most important, pions are shown separately in Fig.~\ref{Tswitch}(b).
Thus, a pion chemical potential of nearly
$80\,\mbox{MeV}$ may be acquired. Similar values 
have been found previously \cite{RappWambach,Grassi-Mu}. See \cite{Pratt-Chemical} for
a discussion of how inelastic reactions can reduce these values.

\begin{figure}[tbp]
\begin{center}
\includegraphics[height=3.0in,width=3.0in]{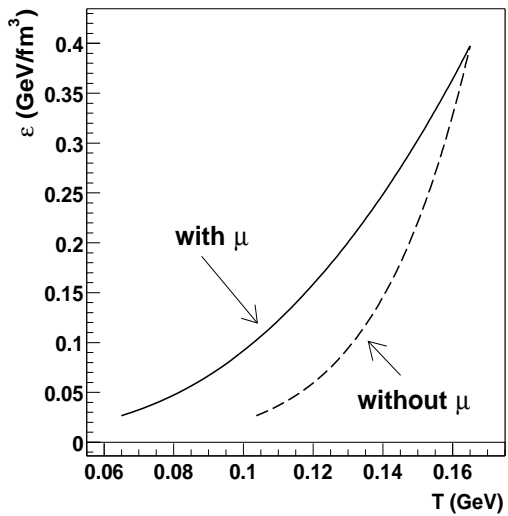}
\end{center}
\caption[The energy density as a function of temperature with
and without chemical freezeout]{
\label{Ep}
The energy density as a function of temperature with and
without chemical freezeout for the SPS ($s/n_{B}$=42).
}
\end{figure}

\begin{figure}[tb!p]
\begin{center}
\includegraphics[height=2.8in,width=2.8in]{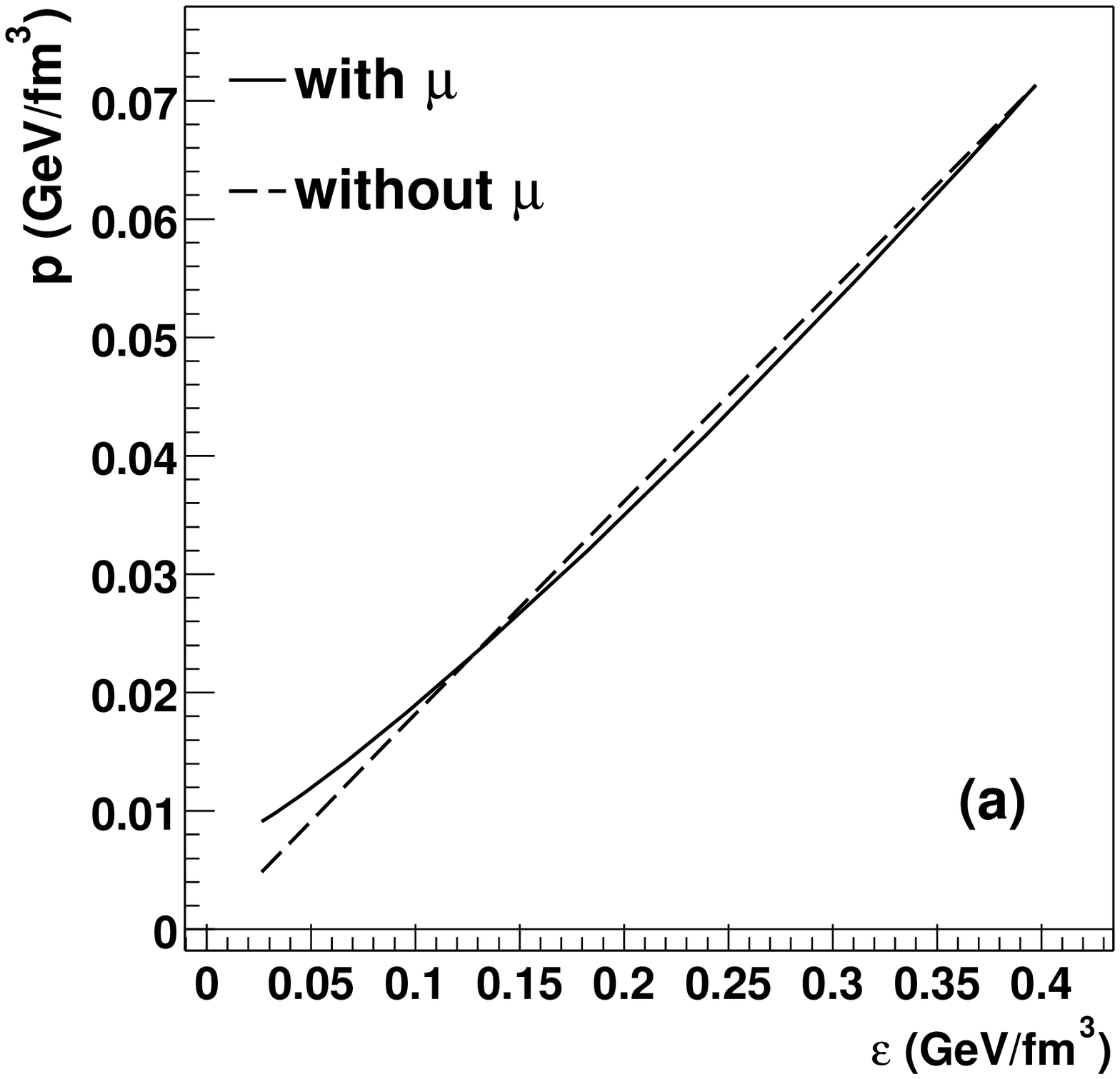}
\hspace{-0.4in}
\includegraphics[height=2.8in,width=2.8in]{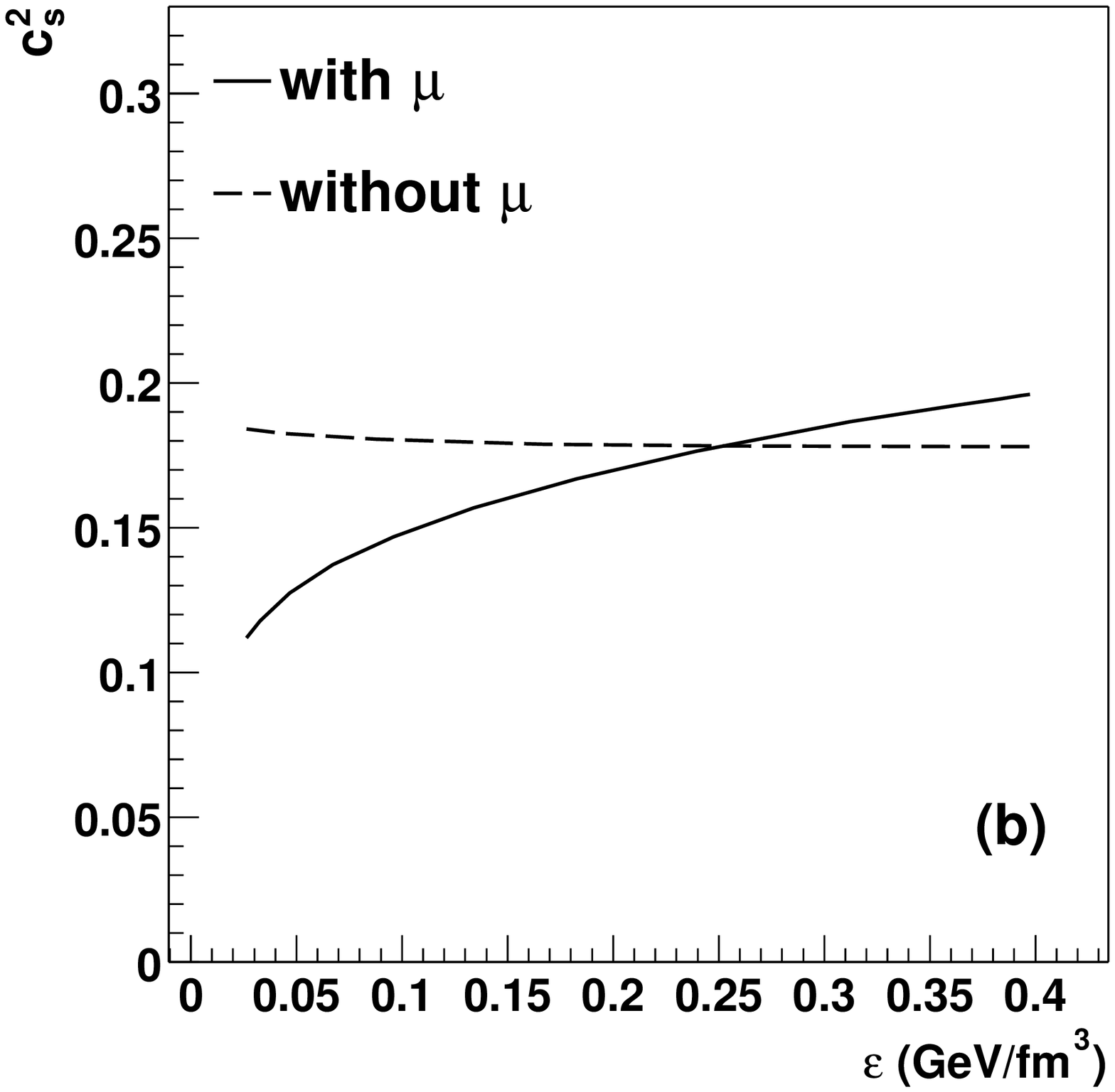}
\includegraphics[height=2.8in,width=2.8in]{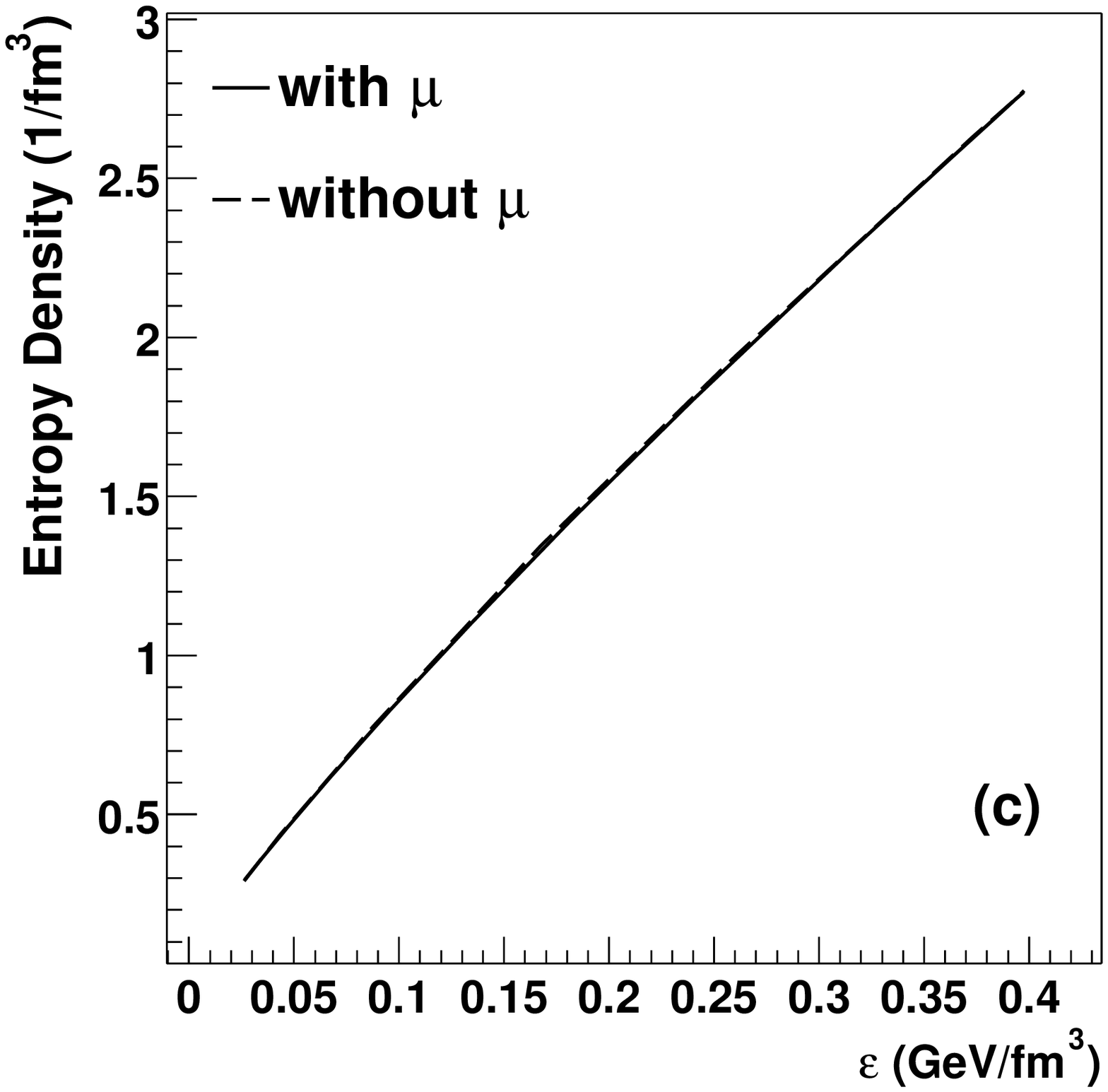}
\end{center}
\caption[The pressure, entropy density, and sound speed as functions
of energy density with and without chemical freezeout.]{
\label{pvsemu}
The (a) pressure, (b) sound speed squared and (c) entropy density
as functions of the energy density with and without chemical freezeout
at the SPS ($s/n_{B}$=42). The analogous curves at RHIC are only slightly different.
}
\end{figure}

Now we return to the EOS. The energy density
as a function of temperature, with and without
the chemical potentials, is shown in
Fig.~\ref{Ep}. The principal observation is  
%%%%%%%%%%%%%%%%%%%%%%%%%%%%%%%%%%%%%%%%%%%%%%%%%%%%%%%%%%%%%%%%%%%%%%%%
%%%%%%%%%%%%%%%%%%%%%%%%%%%%%%%%%%%%%%%%%%%%%%%%%%%%%%%%%%%%%%
clear: without chemical freezeout the energy density
drops very rapidly as a function of temperature, since
the total number of particles drops.
For later discussion, the temperature is tabulated as
a function of energy density with and without chemical
potentials, in Table~\ref{ttemperature}.
%%%%%%%%%%%%%%%%%%%%%%%%%%%%%%%%%%%%%%%%%%%%%%%%%%%%%%%%%%%%%%%%%%%%%%%%
\begin{table}
\begin{tabular}{|l|r|r|} \hline
Energy Density ($\mbox{GeV/fm}^{3}$) & T(MeV) with $\mu_{\pi}>0$ & T(MeV) with
$\mu_{\pi}=0$ \\ \hline\hline
0.364 & 160 & 163 \\ \hline
0.249 & 140 & 153 \\ \hline
0.158 & 120 & 142 \\ \hline
0.122 & 110 & 135 \\ \hline
0.091 & 100 & 129 \\ \hline
0.067 & 90 & 122 \\ \hline
0.047 & 80 & 115 \\ \hline
\end{tabular}
\caption[Table of temperature versus energy density, with and without
chemical freezeout.]{
\label{ttemperature}
Temperature versus energy density, with and without chemical
freezeout at the SPS ($s/n_{B}$ = 42).
}
\end{table}
However, to find the hydrodynamic solution the
most important quantity is not the relationship between
energy density and temperature, but
the relation between the energy density and pressure -- the EOS.
Fig.~\ref{pvsemu} shows the pressure,
%%%%%%%%%%%%%%%%%%%%%%%%%%%%%%%%%%%%%%%%%%%%%%%%%%%%%%%%%%%%%%%%%%%%%%%%
%%%%%%%%%%%%%%%%%%%%%%%%%%%%%%%%%%%%%%%%%%%%%%%%%%%%%%%%%%%%%
the squared sound speed $c_{s}^2$, and the entropy density versus the
energy density along the
adiabatic path. Along the adiabatic path,
these quantities are all related because
\begin{eqnarray}
\left(\frac{dp}{de}\right)_{\left\{n_{i}/s\right\}} &\equiv& \,c_{s}^{2} \\
\left(\frac{ds}{de}\right)_{\left\{n_{i}/s\right\}} &=& \frac{s}{p+e} .
\end{eqnarray}
(The second relation  follows by noting that
$\left(\frac{ds}{de}\right)_{n_i/s} = 
\left( \frac{ds}{de} \right)_{n_i} + 
\left( \frac{ds}{dn_i}\right)_{e} \frac{n_i}{s} 
\left(\frac{ds}{de}\right)_{n_i/s}$, solving for 
$\left(\frac{ds}{de}\right)_{n_i/s}$ and using thermodynamic
identities). 
Unlike energy density versus temperature, pressure
is scarcely modified by the chemical potentials. Differentiating
the pressure, we see that
$c_{s}^{2}$ is reduced somewhat at low temperatures when
chemical freezeout is incorporated. Integrating to find the
entropy, we see that the entropy as a function of energy
is nearly identical with and without chemical freezeout.

\begin{figure}[tb!p]
\begin{center}
\includegraphics[height=3.0in,width=3.0in]{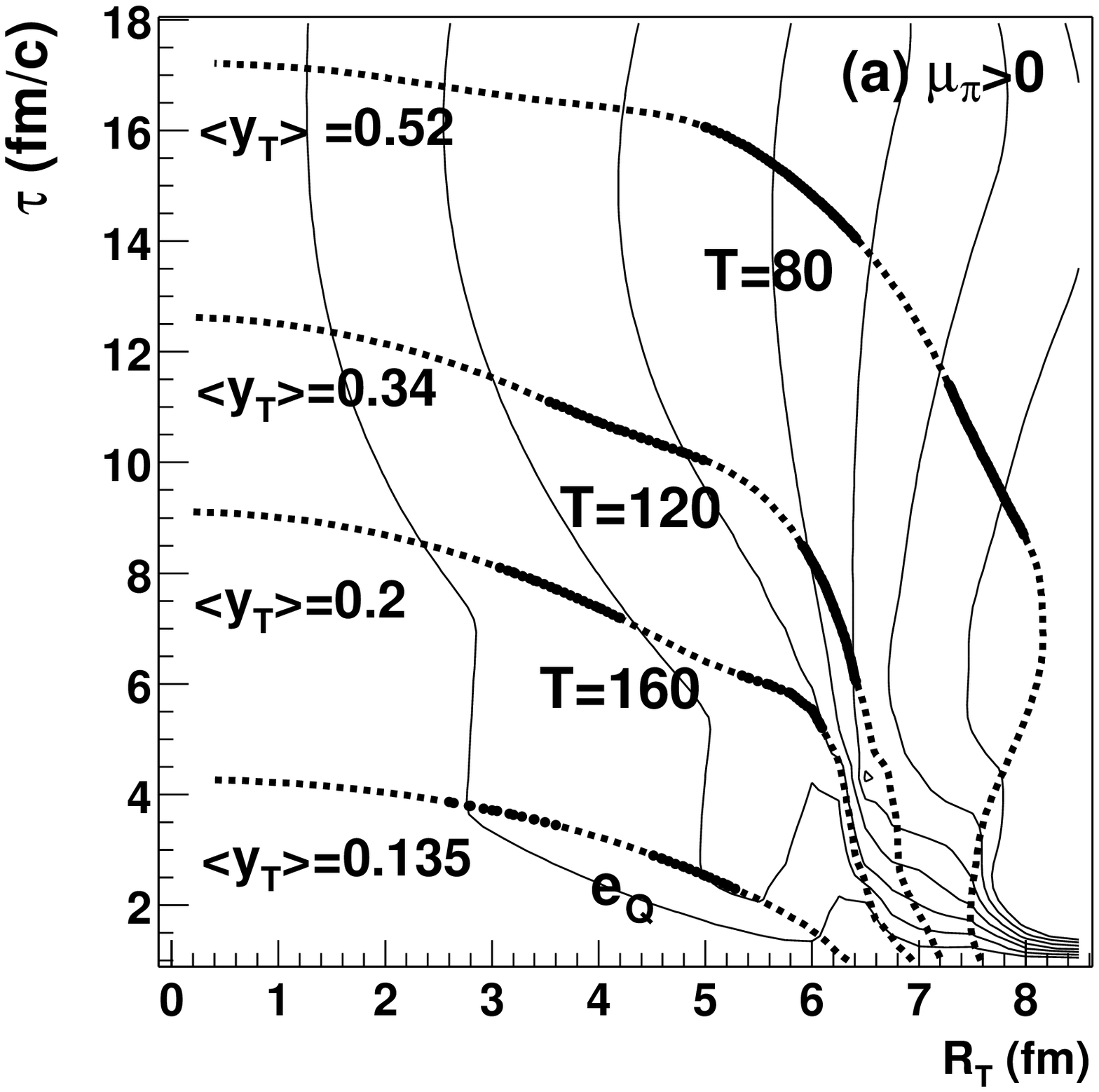}
\vspace{-.2in}
\includegraphics[height=3.0in,width=3.0in]{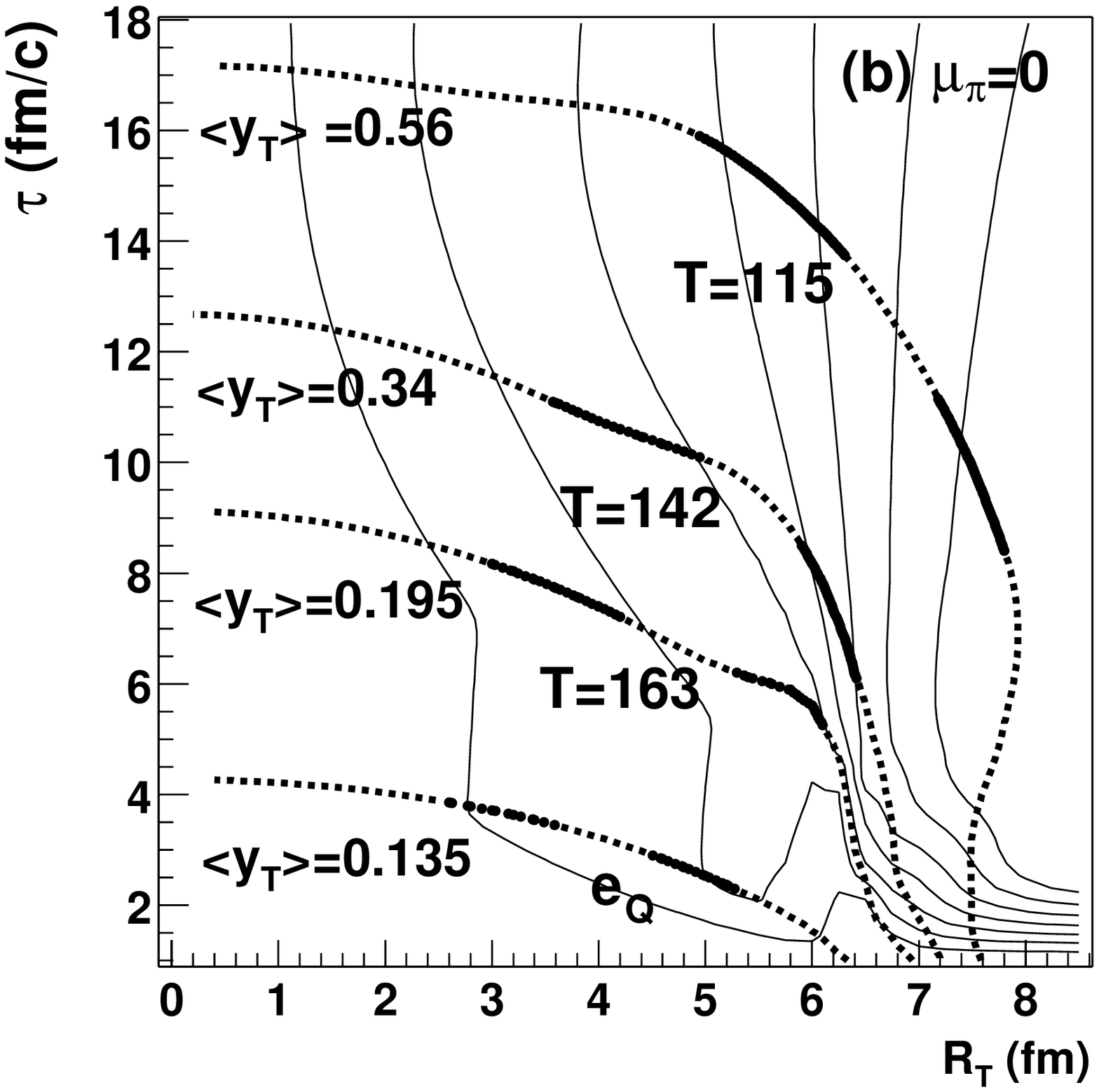}
\end{center}
\caption[The hydrodynamic solution with and
without chemical
freezeout.] { \label{HydroMu} The hydrodynamic solution (a) with and (b) without
chemical freezeout at the 
SPS (PbPb, $\sqrt{s}$=17 GeV A, b=0 fm, $s/n_{B}$=42). The thick arcs show
contours of constant energy density. The first contour indicates
the start of the mixed phase, $e_{Q}$. The next three contours indicate
energy densities corresponding to temperatures (a) T = 160, 120, 80\,MeV
with
chemical freezeout and (b) T=163, 142 , 115 \,MeV without chemical
freezeout, see
Table~\ref{ttemperature}.
The thin lines shows contours of constant transverse
fluid rapidity, $y_{T}=0.1, 0.2, \dots, 0.7$ .
Walking along the thick
arcs, the arc is divided into solid and dotted segments. 20\% of the
total entropy passing through the arc passes through each segment. 
$\left\langle y_{T} \right\rangle$ denotes the mean transverse
rapidity (weighted with entropy) on the arc.
}
\end{figure}

From the point of view of dynamics,
this means that the hydrodynamic solutions, with and without chemical
freezeout,
are nearly the same when expressed as a function
of energy density. The flow velocities on
a freezeout surface of constant energy density are independent of
whether or not chemical potentials are included. However, the temperature
on that freezeout surface depends dramatically on chemical freezeout, as
can be seen from Table~\ref{ttemperature}. To make
this point clear, the hydrodynamic solutions are
illustrated in Fig.~\ref{HydroMu} with and without chemical
%%%%%%%%%%%%%%%%%%%%%%%%%%%%%%%%%%%%%%%%%%%%%%%%%%%%%%%%%%%%%%%%%%%%%%%%
%%%%%%%%%%%%%%%%%%%%%%%%%%%%%%%%%%%%%%%%%%%%%%%%%%%%%%%%%%%%%%%%%%%%%%%%
potentials. The solutions are similar; the
small differences can be traced to small differences in the
speed of sound.

Often in hydrodynamic simulations not incorporating
chemical freezeout, the freezeout surface
$T\approx130\,\mbox{MeV}$ is taken \cite{Huovinen,Hirano-Flow}. Using Table~\ref{ttemperature},
this roughly corresponds to a temperature of $T\approx100\,\mbox{MeV}$.
Although this temperature seems low, it is not out of
keeping with the phenomenological freezeout temperatures
extracted from thermal fits to radial \cite{SSH-FlowProfile}
and recent elliptic flow data \cite{STAR-EllipticParticle}. The
extent to which this low freezeout temperature is
seen in a hadronic cascade is addressed in the next section.

\section{RQMD and Chemical Freezeout}
\label{RQMD}

Returning to hydro+cascade, we can study the extent to
which chemically hydrodynamics reproduces the dynamic response of
the cascade. An important observation is that
with the assumption of chemical equilibrium,
the hydrodynamics is independent of the dominant reactions in the cascade.
With chemical freezeout, reactions such as
$\pi N\rightarrow \Delta \rightarrow \pi N$, are encoded into
the conservation laws and ultimately into the chemical
potentials which affect the spectra. 
Baryons are accelerated
only by reducing the kinetic energy of the pions. In
a hydrodynamic language, the flow velocity is increased only
by rapidly decreasing the temperature.

\begin{figure}[tb!p]
\begin{center}
\includegraphics[height=2.8in,width=2.8in]{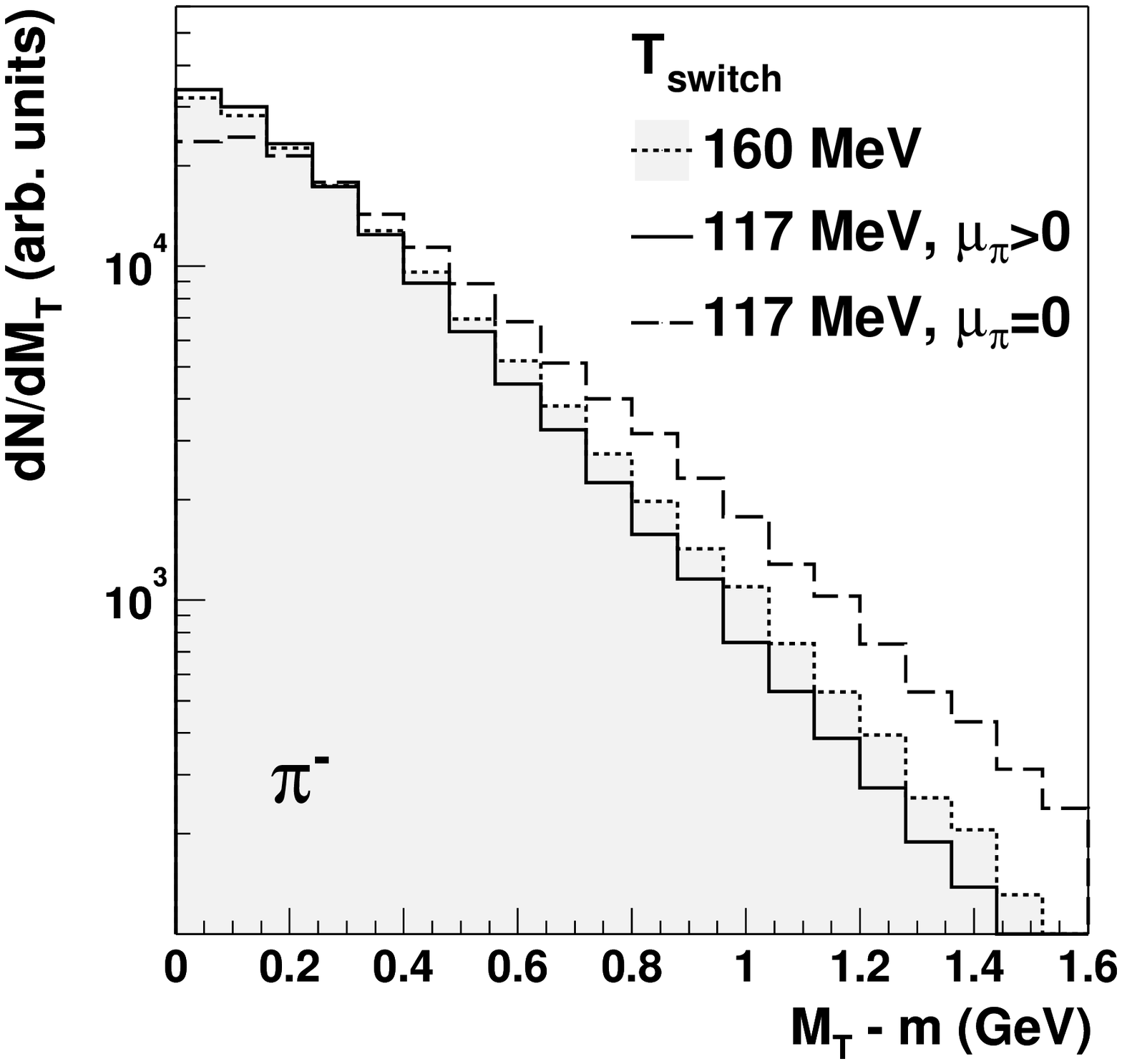}
\hspace{-0.4in}
\includegraphics[height=2.8in,width=2.8in]{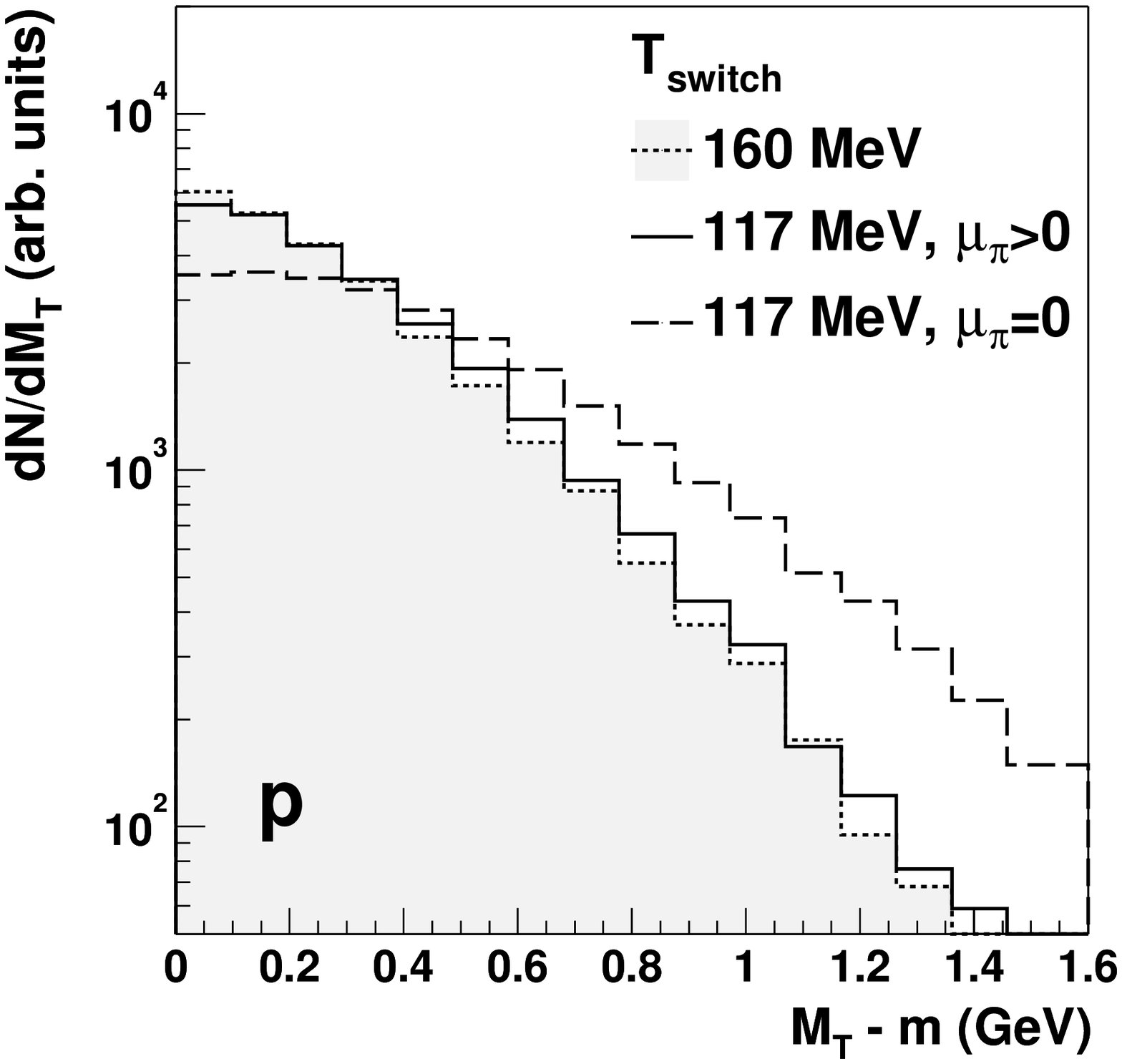}
\hspace{-0.4in}
\includegraphics[height=2.8in,width=2.8in]{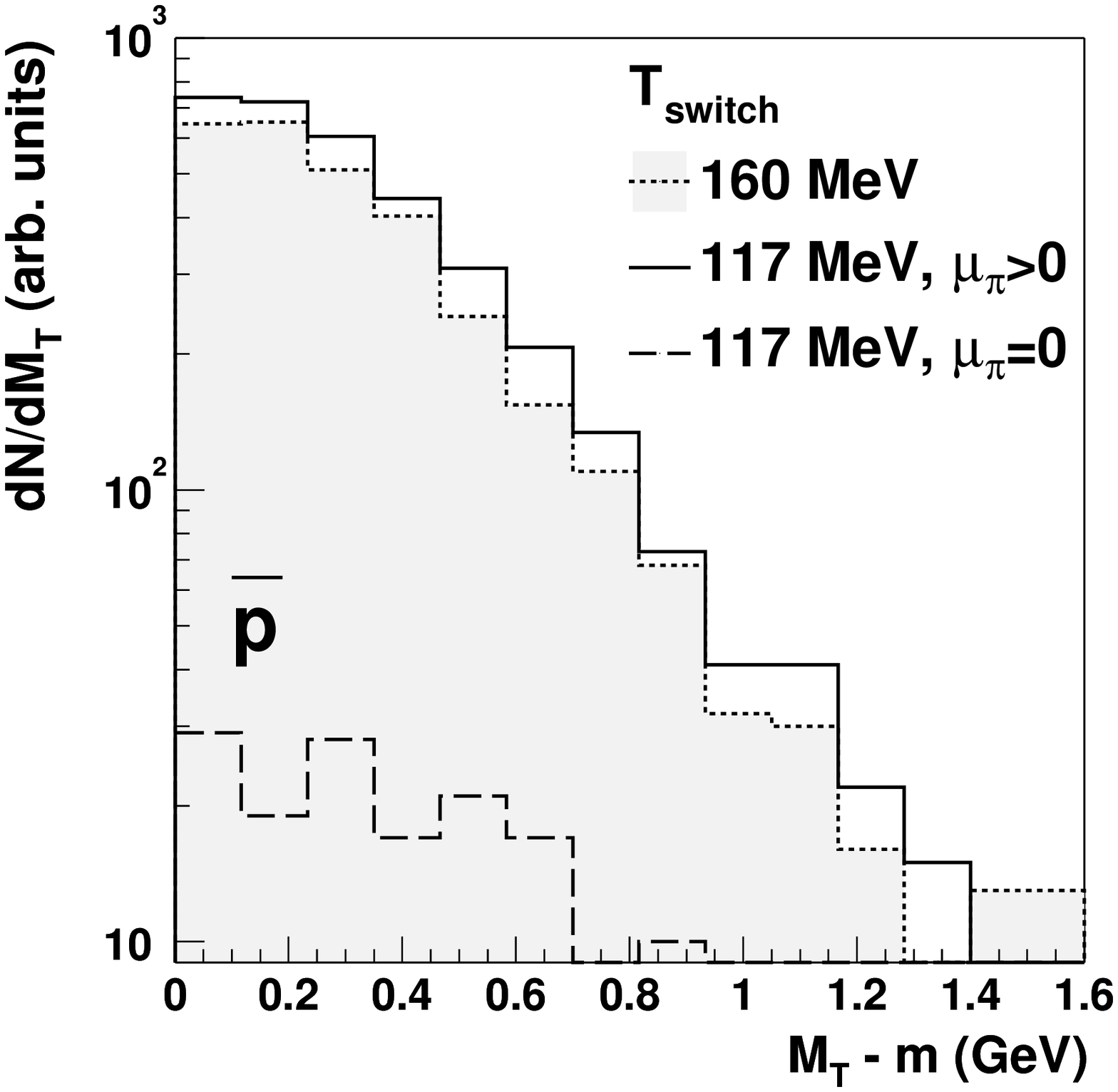}
\end{center}
\caption[Sensitivity of spectra to $T_{switch}$ with and without chemical
freezeout]
{\label{NaiveSpectra}
Sensitivity of particle spectra at the SPS (PbPb, $\sqrt{s}$=17 GeV A, b=6 fm, 
$s/n_{B}$=42) to $T_{switch}$ with chemical freezeout,
$\mu_{\pi}>0$
and without, $\mu_{\pi}=0$. The spectra are for $\pi^{-}$,
$p$, and $\bar{p}$. The
filled histograms are $T_{switch}=160$. The other curves
are for $T_{switch}=117$.
}
\end{figure}

One of the
problems with the hydro+cascade approach \cite{HydroUrqmd}
is the sensitivity to the switching temperature.
To make a smooth transition from hydrodynamics into the cascade,
the same conservation laws have to be implemented in both
approaches.
In Fig.~\ref{NaiveSpectra}, we adjust the switching
%%%%%%%%%%%%%%%%%%%%%%
%%%%%%%%%%%%%%%%%%%%%%%
temperature from $T_{switch}=160\,\mbox{MeV}$ to $T_{switch}=117\,\mbox{MeV}$ , and
compare spectra in which chemical freezeout is incorporated,
$\mu_{\pi}>0$, to spectra in which chemical freezeout is
not incorporated,  $\mu_{\pi}=0$. Ideally, the spectra should be
insensitive
to $T_{switch}$.

First,  notice that if chemical freezeout is not incorporated the yields of
$K$ and $\bar{p}$ are reduced by factors of 
$e^{-\frac{M_K}{T} - \frac{M_K}{T_c}}$ and
$e^{-\frac{M_N}{T} - \frac{M_N}{T_c}}$, 
or 3 and 7 for temperatures of $T\approx120\,\mbox{MeV}$.
Furthermore, the final flow is too strong, since both the
the mass energy and  the kinetic energy are converted into
flow.
Second, notice that once chemical freezeout is incorporated,
the sensitivity to the switching temperature
is small. This is because the freezeout
energy density is higher;  therefore it is ``as if'' the
freezeout temperature were higher.
Close inspection of the pion spectrum shows that
the final spectrum with chemical freezeout is too cool relative to
the Hydro+cascade model. This
rapid cooling will be addressed below.

\begin{figure}[tb!p]
\begin{center}
\includegraphics[height=2.8in,width=2.8in]{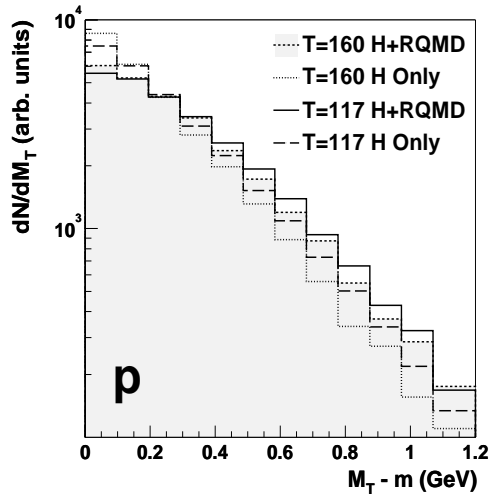}
\end{center}
\caption[The effect of hadronic scattering on spectra
with and without chemical freezeout] {
\label{SpectraScatt}
The effect of hadronic scattering on the proton spectrum at the 
SPS (PbPb, $\sqrt{s}$=17 GeV A, b=6 fm, $s/n_{B}$=42) 
with and without chemical freezeout.
The temperature $T_{switch}$,
where the switch from hydro to cascade made is
given in the figure. H(ydro)+RQMD denotes  the spectra
with subsequent hadronic rescattering in the
cascade. H(ydro) Only denotes a spectrum in
which only resonance decays but no rescattering have been accounted
for.
}
\end{figure}

A naive explanation for the insensitivity to
 $T_{switch}$ is that the decrease in temperature is compensated
by flow.
Fig.~\ref{SpectraScatt} studies this explanation in detail.
%
%%%%%%%%%%%%%%%%%%%%%%%%%%%%%%%%%%%%%%%%%%%%%%%%%%%%%%%%%%%%%%%%%%%%%%%%
%%%%%%%%%%%%%%%%%%%%%%%%%%%%%%%%%%%%%%%%%%%%%%%%%%%%%%%%%%%%%%%%%%%%%%%%
Hadronic rescattering increases
the mean $M_{T}$ of the nucleons while slightly decreasing the mean
$M_{T}$ of the pions. Although these general trends are reproduced
by the hydrodynamics, with the hydrodynamics
the pion cooling is too rapid and
the nucleon acceleration is slightly too large.
These facts
may be gleaned from a very close inspection of the 
four curves in Fig.~\ref{SpectraScatt}. However, the
differences are all small and it is difficult to tell
the difference between chemically frozen hydrodynamics
and free streaming from spectra alone.

\begin{figure}[tb!p]
\begin{center}
\includegraphics[height=3.2in,width=3.2in]{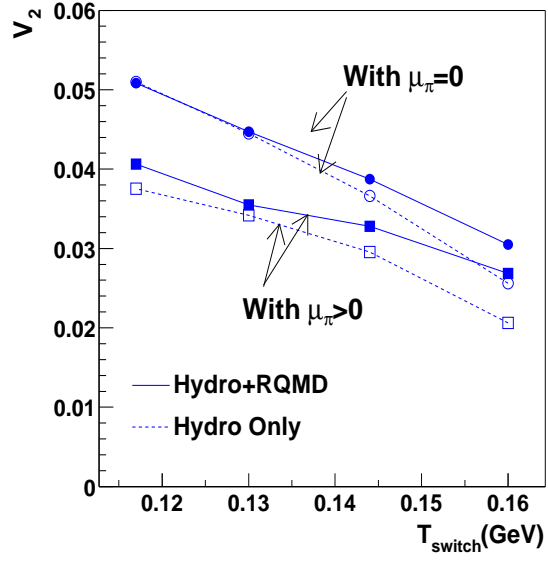}
\end{center}
\caption[The dependence of elliptic flow on $T_{switch}$
with and without chemical freezeout] {
\label{TswitchV2}
$v_{2}$ of pions at the 
SPS (PbPb, $\sqrt{s}$=17 GeV A, b=6 fm, $s/n_{B}$=42) 
as a function of $T_{switch}$ with chemical
freezeout, $\mu_{\pi} > 0$, and without, $\mu_{\pi}=0$
}
\end{figure}

Instead of  spectra, we study  the elliptic flow when
chemical potentials are included. Fig.~\ref{TswitchV2} 
%%%%%%%%%%%%%%%%%%%%%%%%%%%%%%%%%%%%%%%%%%%%%%%%%%%%%%%%%%%%%%%%%%%%%%%%
%%%%%%%%%%%%%%%%%%%%%%%%%%%%%%%%%%%%%%%%%%%%%%%%%%%%%%%%%%%%%%%%%%%%%%%%
shows the sensitivity of elliptic flow to $T_{switch}$.
The elliptic flow  remains fairly  sensitive
to the switching surface and increases by  25\% as
the switching temperature is lowered. The hydrodynamic contribution
increases while the RQMD contribution decreases 
as $T_{switch}$ is lowered. The fact that
the Hydro+RQMD curve is flatter than the Hydro Only curve indicates
that hydro+cascade model is partially successful.

\begin{figure}[tb!p]
\begin{center}
\includegraphics[height=3.2in,width=3.2in]{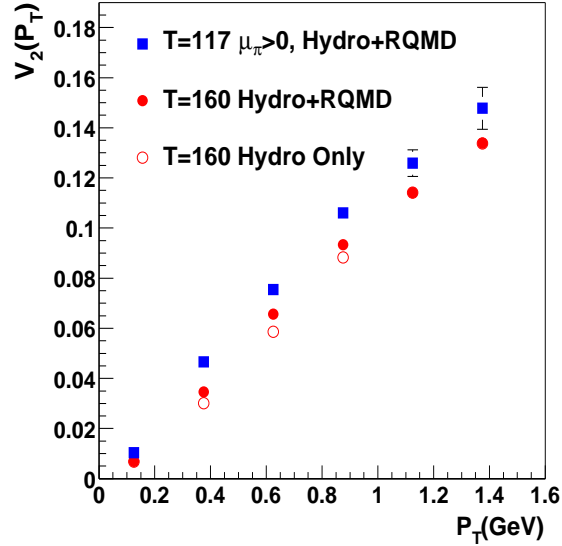}
\end{center}
\caption[The sensitivity of the elliptic flow spectrum to $T_{switch}$
with and without chemical freezeout] {
\label{MuV2Pt}
The elliptic flow spectrum of  pions at the SPS (PbPb, $\sqrt{s}$=17 GeV A, b=6 fm, $s/n_{B}$=42) 
as a function of $T_{switch}$ with chemical
freezeout $\mu_{\pi} > 0$, and without, $\mu_{\pi}=0$.
}
\end{figure}

Additional information is gained with Fig.~\ref{MuV2Pt},  which  examines
%%%%%%%%%%%%%%%%%%%%%%%%%%%%%%%%%%%%%%%%%%%%%%%%%%%%%%%%%%%%%%%%%%%%%%%%
%%%%%%%%%%%%%%%%%%%%%%%%%%%%%%%%%%%%%%%%%%%%%%%%%%%%%%%%%%%%%%%%%%%%%%%%
the elliptic flow spectrum, $v_{2}(p_{T})$.
The rapid cooling of the hydrodynamic response with chemical
freezeout (T=117 $\mu_{\pi}>0$, Hydro+RQMD)
is seen as an increase in elliptic flow at modest $p_{T}$,
relative to the normal curves (T=160 Hydro+RQMD).
By comparing the $T=160\,\mbox{MeV}$ Hydro Only and Hydro+RQMD
points, we see that in the cascade the pions cool only slightly.
Nevertheless, the general trend is the same for both  the hydro with
chemical freezeout and the cascade with chemical freezeout.

%The pion
%slope with $T_{switch}\approx117$
%is smaller than the pion slope when $T_{switch}\approx160$. For
%the nucleons, hadronic rescattering after $T_{switch}\approx160$
%substantially increases the $\langle M_T \rangle$ (compare T=160 H Only
%and
%T=160 H+RQMD). When the switching temperature
%is decreased to $T_{switch}=117$, the flow velocity increases(compare
%T=160 H Only and T=117 H Only). Subsequently
%hadronic rescattering increases $\langle M_T \rangle$ further, although
%the increase is not as much as for $T_{switch}=160$.

\begin{figure}[tb!p]
\begin{center}
\includegraphics[height=3.2in,width=3.2in]{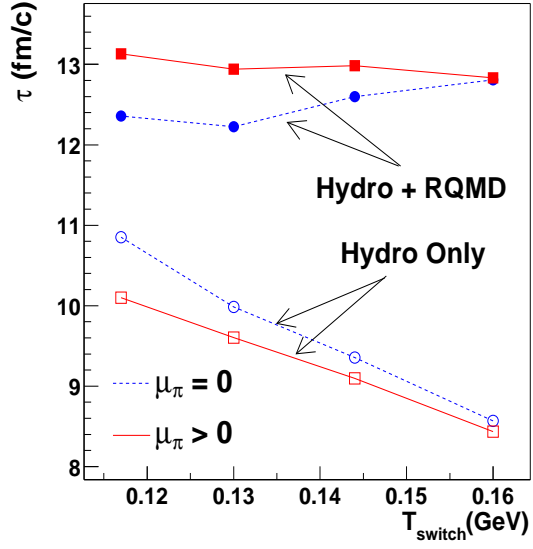}
\end{center}
\caption[The dependence of the mean emission time on
the switching temperature with and without chemical freezeout]
{\label{Time}
The mean emission time $<\tau>$ at the SPS (PbPb, $\sqrt{s}$=17 GeV A, b=6\,fm,  $s/n_{B}$=42) as a function
of $T_{switch}$ with chemical freezeout $\mu_{\pi}>0$,
and without chemical freezeout, $\mu_{\pi}=0$.
}
\end{figure}

We now turn to the distribution of pions at freezeout.
The density of
pions at freezeout is approximately  constant  as a  function of impact
parameter
and beam energy \cite{Htoh}. Now we interpret this density in
thermal terms. In the cascade, the density of pions at freezeout
is the same as the density of pions in a hadron gas with
$T\approx110\,\mbox{MeV}$ and $\mu_{\pi}\approx 70\,\mbox{MeV}$.
Notice Fig.~\ref{Time}, which shows the mean emission time
%%%%%%%%%%%%%%%%%%%%%%%%%%%%%%%%%%%%%%%%%%%%%%%%%%%%%%%%%%%%%%%%%%%%%%%%
%%%%%%%%%%%%%%%%%%%%%%%%%%%%%%%%%%%%%%%%%%%%%%%%%%%%%%%%%%%%%%%%%%%%%%%%
as the switching temperature is lowered. As with the
spectra, the mean emission time after cascading is insensitive to the
switching temperature. The system of hadrons introduced
into the cascade expands until the density reaches a
certain value and subsequently breaks up.
Since the number of hadrons is conserved during
the hydrodynamic evolution, the final breakup does not
depend on where the hydrodynamics is stopped and where the
cascade begins.

\begin{figure}[tb!p]
\begin{center}
\includegraphics[height=3.2in,width=3.2in]{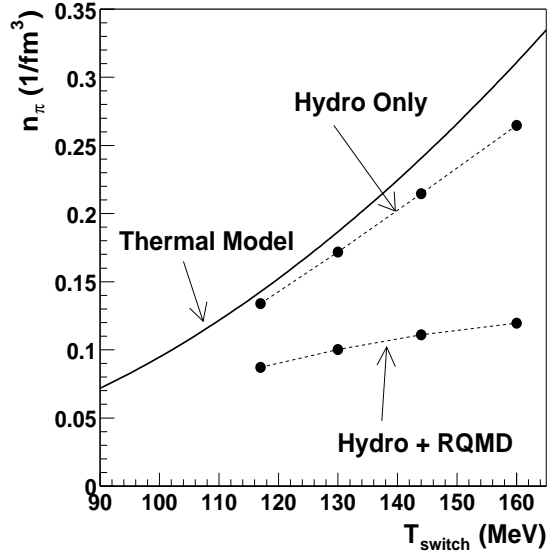}
\end{center}
\caption[The hadron density in RQMD at freezeout]
{
\label{FigDensity}
The pion density at freezeout in RQMD (see Eq.~\ref{EqDensity})
as a function of $T_{switch}$ at the 
SPS (PbPb, $\sqrt{s}$=17 GeV A, b=6 fm, $s/n_{B}$=42).
The Thermal Model curves indicate the density
of pion number (see Eq.~\ref{ThermalDensity})
in a hadron gas, i.e. $n_{\pi} + 2 n_{\rho} + ...$ .
}
\end{figure}

We now try to measure the RQMD pion density at freezeout.
The cascade does not have a freezeout surface, as is normally assumed
in an idealized hydro picture. Rather, particles are emitted
per unit time and volume, i.e. $\frac{dN}{\tau d\eta \, dx\,dy\,d\tau}$.
We define the freezeout density in the transverse plane as
\begin{eqnarray}
\label{EqDensity}
n_{\pi}^{RQMD} &\equiv& \frac{1} { \pi R_{o}^{2} }
\int_{\sqrt{x^2 + y^2} < R_{o}} dx\,dy\, d\tau\,\frac{dN}{\tau
d\eta\,dx\,dy\,d\tau},
\end{eqnarray}
where $R_{o}$ is taken to be 3\,fm.
This density, $n_{\pi}^{RQMD}$, with
and without hadronic rescattering is shown in Fig.~\ref{FigDensity}.
%%%%%%%%%%%%%%%%%%%%%%%%%%%%%%%%%%%%%%%%%%%%%%%%%%%%%%%%%%%%%%%%%%%%%%%%
%%%%%%%%%%%%%%%%%%%%%%%%%%%%%%%%%%%%%%%%%%%%%%%%%%%%%%%%%%%%%%%%%%%%%%%%
Of course, without rescattering, the density of pions should
reflect the density  of pions on the freezeout surface. With
the EOS used in this work, this number density is simply the
zero-th component of $J_{H_{\pi}}^{\mu}$ or
\begin{eqnarray}
\label{ThermalDensity}
J_{H_{\pi}}^{0}= n_{\pi} + 2 n_{\rho} + n_\Delta\dots .
\end{eqnarray}
This thermal density of pions is also shown in
Fig.~\ref{FigDensity} and gives a reasonable description
of the Hydro Only points. When rescattering is included,
the density decreases until a constant value
$\approx 0.1\,\mbox{fm}^{3}$, is reached.
Comparing the thermal model curves to the Hydro+RQMD points,
we see that the freezeout densities are equal for
$T\approx100-110\,\mbox{MeV}$ and $\mu_{\pi}\approx80\,\mbox{MeV}$.
Although this temperature is low, it is not out of
keeping with the numbers extracted from hydrodynamic fits to
the data \cite{STAR-EllipticParticle,SSH-FlowProfile}.

%%%%%%%%%%%%%%%%%%%%%%%%%%%%%%%%%%%%%%%%%%%%%%%%%%%%%%%%%%%%%%%%%%%%%%%%

\section{Summary of Chemical Freezeout}

We have constructed an EOS consistent with
chemical freezeout. The relationship between
pressure and energy density is approximately
the same as the standard EOS. However,
these two EOS exhibit dramatically different
relationships between energy density and temperature. 
At a practical level, this means that the results from
a hydrodynamic calculation without chemical freezeout may
be consistently converted to a calculation with chemical freezeout by
simultaneously adjusting the yields and the temperature
to keep the energy density constant. Table ~\ref{ttemperature} gives
the required conversion factors.

We used  this chemically consistent EOS to study the
hydrodynamic + hadronic cascade model, H2H \cite{Htoh}. Since,
chemical freezeout is incorporated via the hadronic
cross sections into the cascade,
the same conservation laws are implemented
in both the hydrodynamics and the cascade. With this congruence, 
we examined the model sensitivity
to the switching temperature from hydro to cascade,  $T_{switch}$.
The spectra are insensitive to the switching temperature for
$120\,\mbox{MeV} < T_{switch} < 160\,\mbox{MeV}$. On the other hand,
elliptic flow at the SPS remains mildly sensitive to $T_{switch}$ even
when chemical freezeout is incorporated. The insensitivity
of the model's results to $T_{switch}$ 
partially validates the
hydro+cascade approach.

An important feature of a chemically frozen EOS
is that the principal reactions in the cascade,
e.g. $\pi N\rightarrow \Delta \rightarrow \pi N$,
are encoded into the hydrodynamics through the
chemical potentials. Thus, in the cascade,
the nucleons are accelerated only by reducing the kinetic
energy of the pions.
With a chemically frozen EOS,
this qualitative feature is reproduced by the hydrodynamics.
However, as might be expected,
the cooling and attendant acceleration are larger in the hydrodynamics
than in the cascade. When the cascade is replaced
with the hydrodynamics, the pion spectrum is slightly
steeper while the nucleon spectrum is slightly flatter.
Pion cooling
may also be seen with $v_{2}(p_{T})$. Here again, the small
changes in $v_{2}(p_T)$ due to hadronic rescattering
are strongly magnified by the hydrodynamics with chemical freezeout.

By incorporating chemical freezeout, many of the
qualitative features of the cascade's evolution are reproduced with the
hydrodynamics (see \cite{Pasi-Photons} for further investigations of this
point). Thus, a thermal interpretation of the cascade's
response is partially justified. Within RQMD, we find
the density
of pions  at freezeout to be $n_{\pi}\approx 0.12\,\mbox{fm}^{-3}$, which
corresponds to $T \approx 110 \,\mbox{MeV}$ and
$\mu_{\pi}\approx70\,\mbox{MeV}$ in thermal terms. Precisely these parameters have
been extracted from thermal fits to hadronic data.
Therefore, 
an understanding  of  the properties of a chemically frozen
hadronic gas  helps to bridge the cascade and thermal descriptions of
the final stages of the heavy ion collision.

\section{Acknowledgments}
I thank Edward Shuryak for continued support.

\end{document}